# Location qubits in a multi-quantum-dot system

Dayang Li[1], Nika Akopian[1]*

[1] DTU Department of Photonics Engineering, Technical University of Denmark; Kongens Lyngby, 2800, Denmark

*Corresponding author. Email: nikaak@fotonik.dtu.dk

**A physical platform for nodes of the envisioned quantum internet is long-sought(*1*). Here we propose such a platform, along with a conceptually simple and experimentally uncomplicated quantum information processing scheme, realized in a system of multiple crystal-phase quantum dots(*2*). We introduce novel location qubits, describe a method to construct a universal set of all-optical quantum gates, and simulate their performance in realistic structures, including decoherence sources. Our results show that location qubits can maintain coherence 5 orders of magnitude longer than single-qubit operation time, and single-qubit gate errors do not exceed 0.01%. Our scheme paves a clear way towards constructing multi-qubit solid-state quantum systems with a built-in photonic interface, such as a multi-qubit quantum register — a key building block of the forthcoming quantum internet.**

Physical platforms where multi-qubit quantum information can be fault-tolerantly processed and faithfully stored, reliably received and sent over long distances have been under intensive study(*1*). The search is active in a multitude of systems including but not limited to superconducting circuits(*3*), trapped ions(*4*), NV centers in diamond(*5*), optomechanical systems(*6*) and various types of quantum dots (*7–10*). A practical platform should possess several essential characteristics: (i) it has to be designable — such that an individual qubit can be designed and built as desired; (ii) it has to be scalable to multiple qubits; (iii) it should have a photonic interface for long-distance communication; and, (iv) it should support the nanoscale footprint of devices — such that a large number of quantum systems can be realistically created. A novel system — crystal-phase quantum dots(*2, 10–12*) — stands out here, as, unlike other platforms, it possesses all four characteristics simultaneously. Its designability is particularly remarkable — it allows for an accuracy of a single atomic layer(*13–15*) during device fabrication.

Here we propose a designable solid-state multi-qubit platform with a photonic interface, exploiting the unique advantages of crystal-phase quantum dots. In these structures, electrons and holes are confined in different spatial regions of a III-V semiconductor nanowire, formed by alternating wurtzite (WZ) and zinc blende (ZB) crystal phases (*2, 12, 16*), as we show in Fig. 1A. Consequently, excitons are spatially indirect, but can share a single charge state — a hole, for instance. Such a hole can, therefore, act as a link (*10*), connecting spatially separated electron states, which we exploit in this work (likewise, an electron can connect two spatially separated hole states). An InP nanowire is chosen for numerical studies in this work, but in principle any quantum confinement structure with type-II band alignment is a suitable platform for the quantum information processing scheme to be introduced.

On a neighbouring pair of crystal-phase quantum dots, an electron can be optically loaded(*10*) either into the left or the right quantum dot, defining its location states $|0\rangle$ or $|1\rangle$, respectively (Fig. 1A). Both location states are connected to a common spatially separated excited state $|X^-\rangle$, forming a three level $\Lambda$-type system ($\Lambda$-system) (Fig. 1B). In such a $\Lambda$-system, we can coherently manipulate the location states using optical pulses(*17*). We now define a location qubit as a linear combination of $|0\rangle$ and $|1\rangle$. In the following sections we show how to implement single- and two-qubit gates on the location qubits, and consequently a universal set of all-optical quantum gates.

We note that location qubits are similar to charge qubits (*7*, *9*), which also encode quantum information in charge states. The location qubit, however, emphasizes on the localization of an individual electron (or hole) on two individual (and not coupled) quantum dots. The stable localization is ensured by a sufficient inter-dot separation (see Supplementary Information B.1 (*18*)). We initialize, manipulate, and measure location qubits optically, unlike charge qubits that are typically operated electrically (*19*, *20*). Electrical operation is identified as a substantial source of charge noise (*19–21*) and, as a consequence, short coherence times (*9*, *19*, *20*) of charge qubits. Such charge noise is not present in our location qubits, leading to significantly longer qubit coherence times.

We start by introducing a scheme to implement an arbitrary single-qubit gate on a single location qubit, using a Bloch sphere. A single-qubit state $|\psi_i\rangle$ is represented by its Bloch vector with coordinates $\mathbf{S} = (\langle\sigma_x\rangle, \langle\sigma_y\rangle, \langle\sigma_z\rangle)$. A single-qubit gate acting on the state $|\psi_i\rangle$ is a rotation of its Bloch vector with respect to a specific rotation axis for a certain rotation angle. To define the appropriate axis and angle, we study the dynamics of our system. We adopt the driven $\Lambda$-system Hamiltonian under the rotating wave approximation(*22*).

$$H = \begin{bmatrix} 0 & \Omega_0(t)\exp(i\alpha) & 0 \\ \Omega_0(t)\exp(-i\alpha) & \Delta & \Omega_1(t) \\ 0 & \Omega_1(t) & 0 \end{bmatrix}.$$

$\Omega_{0/1}(t) = \Omega_{0/1}\exp(-t^2/(2\sigma_t^2))$ is the real Rabi frequency driving the transition $|0\rangle/|1\rangle \leftrightarrow |X^-\rangle$ with $\sigma_t$ characterizing the driving pulse duration. $\alpha$ is the phase difference between the driving fields and $\Delta$ is the single-photon detuning as illustrated in Fig. 1B.

A qubit manipulation scheme in a $\Lambda$-system has been proposed for a spin system (*23*) and generalized to an arbitrary system (*17*). We apply the scheme to our system of location qubits. On the Bloch sphere, the phase difference $\alpha$ between the Rabi frequencies is precisely the azimuthal angle of the rotation axis. The angle $\beta$ (half of the polar angle $2\beta$) is defined as $\tan(\beta) = \Omega_1(t)/\Omega_0(t)$. The rotation angle $\gamma$ with respect to the rotation axis is a function of the single photon detuning $\Delta$.

$$\gamma(\Delta) = \int_{t_i}^{t_f} \left[\sqrt{\Omega_0(t)^2 + \Omega_1(t)^2 + (\Delta/2)^2} - \Delta/2\right] \mathrm{d}t,$$

with $t_i$ and $t_f$ being a time before and after the application of the driving pulses respectively. We illustrate the single qubit rotation in Figure 1C. Details are available in the supplementary information (*18*) section A. Since $\alpha$ and $\beta$ can be chosen over the entire Bloch sphere, and $\gamma$ can take on any value between 0 and $\pi$, an arbitrary single-qubit rotation can be realized. We simulate the population and fidelity evolution for some representative single-qubit gates (parameters in Figure 1D) and present the results in Figure 2. AThe single qubit gate fidelity is measured by the Jozsa-fidelity(*24*)

$$F(\rho_1, \rho_2) = \left( \text{Tr} \sqrt{\sqrt{\rho_1} \rho_2 \sqrt{\rho_1}} \right)^2,$$

which compares the "likelihood" between two density matrices. Here $\rho_1$ represents the density operator of the system prepared by our scheme, and $\rho_2$ the theoretically ideal final state.

For quantum information processing a universal set of gates is required, a common set(*25*) consists of the Hadamard gate $H$, the phase-T gate $T$ and the two qubit controlled-NOT gate $CNOT$. A two-qubit CNOT gate is required to entangle pairs of qubits. We use an additional quantum dot — acting as a control site between two neighbouring location qubits — two neighbouring triple-quantum-dots to demonstrate a CNOT gate (Fig. 3A) in our system. Here we exploit Coulomb interaction to achieve the control mechanism. An electron confined in a crystal-phase quantum dot exerts a Coulomb potential on the neighbouring quantum dots. The exerted potential is most prominent on the nearest neighbours leading to an energy shift of hundreds of GHz(*26*) but diminishes quickly one quantum dot further away — (less than 1GHz for an interdot separation of more than 25 nm)(*26*) for an interdot separation of more than 25 nm. For a driving field with a Gaussian pulse envelope in the transform limit, the induced energy shift on neighbouring dots will be resolved for a pulse duration (FWHM) longer than 4ps. Our CNOT gate is broken down into three steps. (Fig. 3A):

- **Step 1. Enable the control mechanism.** Any electron located at $|1_c\rangle$ is moved to the control site $|2_c\rangle$ by a Pauli-X gate on the $\Lambda$-system formed between $|1_c\rangle$ and $|2_c\rangle$.

levels of the states $|0_t\rangle$ and $|X_t^-\rangle$ and related transition energies.

- **Step 2. Perform the NOT operation.** This is Pauli-X gate acting on the target qubit ($\Lambda$-system formed between $|0_t\rangle$ and $|1_t\rangle$). The laser frequencies are tuned according to the modified transition energies levels of the target qubit. The charge configurations that does not have an electron at the control site $|2_c\rangle$ will not be affected by this Pauli-X gate.

- **Step 3. Disable the control mechanism.** Restoring the charge configuration of the control qubit by moving the electron at the control site $|2_c\rangle$ back to $|1_c\rangle$.

Our implementation of the CNOT gate is essentially three consecutive Pauli-X gates. As a demonstration, CNOT gate is applied on the state $|1_c 0_t\rangle$ (Fig. 3B). The full population is transferred during the first, second and third step to $|2_c 0_t\rangle$, $|2_c 1_t\rangle$ and $|1_c 1_t\rangle$ respectively. A realistic CNOT gate fidelity close to 99.99% is obtained in our simulation. Examples of CNOT on more general initial states are available in the supplementary information C (*18*).

In our schemes for single- and two-qubit gates, the spin degree of freedom is not an information carrier, and the electrons have indeterminate spins. It is well known that optical transitions only happen for particular combinations of electron spin and laser polarization due to selection rules. To ensure the operation, our scheme requires that a matching optical polarization should always be available no matter which spin state the electron is in. This can be ensured by using a depolarizer for the excitation laser to create a statistical mixture of polarizations. Another potential influence is spin-orbit coupling, which mixes the spin and orbital states to form new eigenstates that have neither a definite spin nor orbital, and is expected to be significant for heterostructures and III-V semiconductors. But for sufficiently small quantum dots (in our system, less than 20 nm (*18*)) only a single localized orbital exists, the eigenstates form a degenerate doublet(*27*) in zero magnetic field. Most importantly, they share the same orbital wavefunction. Again a mixture of driving field polarizations solves the problem.

Our location qubits are free from the decoherence mechanisms intrinsic to spin and electrical operations. We identify the main decoherence mechanisms as spontaneous emission from the $|X^-\rangle$ states and dephasing due to electron-phonon interaction(*28*). Spontaneous emission is a Markovian process governed by the spontaneous emission rate. We cover the evaluation of the spontaneous emission rates in the supplementary information B (*18*). We obtain an upper bound of $\gamma_{\text{sp}} = 7.79 \cdot 10^8 \text{s}^{-1}$ for realistic quantum dot dimensions. However since the two eigenstates used for single-qubit gates are purposefully chosen to contain negligibly small $|X^-\rangle$ components, spontaneous emission is only present during gating operations and its effect is fairly limited due to the vast difference in time scale with single qubit operations (10ps). Electron-phonon interaction is another source of decoherence inevitable in most solid state systems. In crystal-phase quantum dots we primarily consider deformation coupling to longitudinal acoustic phonons(*28–30*). We present a detailed analysis of electron-phonon interactions in the supplementary information B.2 (*18*). In short, it results in dephasing of the states $|0\rangle$ and $|1\rangle$. Our numerical results indicate that for a typical quantum dot size of 20nm at the temperature of 4K, the dephasing time is $\tau_{\text{dp}} = 1/\gamma_{\text{dp}} = 6\mu\text{s}$. (Fig. 4B) Considering that the typical single-qubit operation time is on the order of 10ps, the dephasing time is 5 orders of magnitude longer.

Further reduction in operational temperature can significantly suppress dephasing, at 1K and 0.1K the dephasing lifetimes become 2.6ms and $10^6$s. In principle the rates that govern various decoherence processes vary depending on a number of factors such as the size of the quantum dot and temperature. However the spontaneous emission rates are set to the same $\gamma_{\rm sp}$ and the realistic dephasing rates to $\gamma_{\rm dp}$ in a Lindblad master equation used to simulate the dynamics of location qubits and the gate fidelites(*18*).

We have numerically studied the effect of decoherence on single-qubit gate fidelity. Single-qubit gate fidelities are simulated for an arbitrary initial state $|\psi_i(\mu,\nu)\rangle = \cos(\mu)|0\rangle + e^{i\nu}\sin(\mu)|1\rangle$ parametrized by $\mu$ and $\nu$. (Fig. 4, C, D and E) A minimum gate fidelity of 99.9927% for $X$, 99.9928% for $H$ and 99.9983% for $S$ are obtained. Since CNOT consists of three consecutive $X$ gates, we expect the gate infidelity to be on the same order of magnitude. An example is given in Figure 3B.

For a quantum register of multiple location qubits on a single nanowire, each quantum dot can vary from a few to approximately 20 nm in length, with a typical separation of 30-50 nm. Comparing with typical nanowire length of 2 $\mu$m, tens of quantum dots can be built into the nanowire. However it is crucial to design the system such that the optical transitions are resolved with respect to the driving lasers.

In summary, we have introduced a multiqubit photonic device based on semiconductor nanowire crystal-phase quantum dots. The location qubits are defined on a pair of neighbouring quantum dots. A universal set of all-optical quantum gates that is conceptually and experimentally simple is proposed and simulated. Single-qubit gate fidelities are expected to exceed 99.99% even in the presence of the main decoherence mechanisms. The proposed scheme is a promising approach to quantum information processing and quantum networking. In conjunction with the recent advancement in single photon sources based on nanowire quantum dots[31], it is expected that a photonic interface can be incorporated into the system, allowing the quantum register to function as a quantum network node, further strengthening the scalability of our system. Recent progress on the extended Pauli principle[32] indicates that our system can also serve as a testing ground for fundamental multipartite fermionic entanglement.

# Acknowledgement

We thank E. V. Denning for the valuable discussion regarding electron-phonon interactions.

**Funding**: European Research Council ERC Grant Agreement No. 101003378 (NA)

**Author Contributions**:

    Conceptualization: DL, NA

    Data curation: DL

    Formal analysis: DL

    Methodology: DL, NA

    Funding acquisition: NA

    Software: DL

    Supervision: NA

    Visualization: DL, NA

    Writing - original draft: DL

    Writing - review & editing: DL, NA

**Competing interests**: The authors declare no competing interests.

# Supplementary Materials

Materials and Methods

Table S1

Fig S1 – S9

References (32 – 37)

# References

1. S. Wehner, D. Elkouss, R. Hanson, Quantum internet: A vision for the road ahead. *Science*. **362** (2018), doi:10.1126/science.aam9288.

2. N. Akopian, G. Patriarche, L. Liu, J. C. Harmand, V. Zwiller, Crystal phase quantum dots. *Nano Lett.* **10**, 1198–1201 (2010).

3. M. Kjaergaard, M. E. Schwartz, J. Braumüller, P. Krantz, J. I.-J. Wang, S. Gustavsson, W. D. Oliver, Superconducting Qubits: Current State of Play. *Annual Review of Condensed Matter Physics*. **11**, 369–395 (2020).

4. C. Monroe, J. Kim, Scaling the ion trap quantum processor. *Science*. **339**, 1164–1169 (2013).

5. L. Childress, R. Hanson, Diamond NV centers for quantum computing and quantum networks. *MRS Bull.* **38**, 134–138 (2013).

6. A. Wallucks, I. Marinković, B. Hensen, R. Stockill, S. Gröblacher, A quantum memory at telecom wavelengths. *Nat. Phys.* **16**, 772–777 (2020).

7. T. Hayashi, T. Fujisawa, H. D. Cheong, Y. H. Jeong, Y. Hirayama, Coherent manipulation of electronic states in a double quantum dot. *Phys. Rev. Lett.* **91**, 1–4 (2003).

8. R. J. Warburton, Single spins in self-assembled quantum dots. *Nat. Mater.* **12**, 483–493 (2013).

9. D. Kim, D. R. Ward, C. B. Simmons, J. K. Gamble, R. Blume-Kohout, E. Nielsen, D. E. Savage, M. G. Lagally, M. Friesen, S. N. Coppersmith, M. A. Eriksson, Microwave-driven coherent operation of a semiconductor quantum dot charge qubit. *Nat. Nanotechnol.* **10**, 243–247 (2015).

10. J. Hastrup, L. Leandro, N. Akopian, All-optical charging and charge transport in quantum dots. *Sci. Rep.* **10**, 14911 (2020).

11. B. Loitsch, J. Winnerl, G. Grimaldi, J. Wierzbowski, D. Rudolph, S. Morkötter, M. Döblinger, G. Abstreiter, G. Koblmüller, J. J. Finley, Crystal phase quantum dots in the ultrathin core of GaAs-AlGaAs core-shell nanowires. *Nano Lett.* **15**, 7544–7551 (2015).

12. M. Bouwes Bavinck, K. D. Jöns, M. Zieliński, G. Patriarche, J. C. Harmand, N. Akopian, V. Zwiller, Photon Cascade from a Single Crystal Phase Nanowire Quantum Dot. *Nano Lett.* **16**, 1081–1085 (2016).

13. S. Lehmann, J. Wallentin, D. Jacobsson, K. Deppert, K. A. Dick, A general approach for sharp crystal phase switching in InAs, GaAs, InP, and GaP nanowires using only group v flow. *Nano Lett.* **13**, 4099–4105 (2013).

14. D. Jacobsson, F. Panciera, J. Tersoff, M. C. Reuter, S. Lehmann, S. Hofmann, K. A. Dick, F. M. Ross, Interface dynamics and crystal phase switching in GaAs nanowires. *Nature*. **531**, 317–322 (2016).

15. J. C. Harmand, G. Patriarche, F. Glas, F. Panciera, I. Florea, J. L. Maurice, L. Travers, Y. Ollivier, Atomic Step Flow on a Nanofacet. *Phys. Rev. Lett.* **121**, 166101 (2018).

16. K. A. Dick, C. Thelander, L. Samuelson, P. Caroff, Crystal phase engineering in single InAs

nanowires. *Nano Lett.* **10**, 3494–3499 (2010).

17. X. Caillet, C. Simon, Precision of single-qubit gates based on Raman transitions. *Eur. Phys. J. D.* **42**, 341–348 (2007).

18. D. Li, N. Akopian, Supplementary information - Universal set of all-optical quantum gates realized in a multi-quantum-dot system.

19. T. Fujisawa, T. Hayashi, S. Sasaki, Time-dependent single-electron transport through quantum dots. *Rep. Prog. Phys.* **69**, 759–796 (2006).

20. B.-C. Wang, B.-B. Chen, G. Cao, H.-O. Li, M. Xiao, G.-P. Guo, Coherent control and charge echo in a GaAs charge qubit. *EPL*. **117**, 57006 (2017).

21. T. Fujisawa, Y. Hirayama, Charge noise analysis of an AlGaAs/GaAs quantum dot using transmission-type radio-frequency single-electron transistor technique. *Appl. Phys. Lett.* **77**, 543–545 (2000).

22. G. S. Agarwal, *Quantum Optics* (Cambridge University Press, Cambridge, 2012).

23. P. Chen, C. Piermarocchi, L. J. Sham, D. Gammon, D. G. Steel, Theory of quantum optical control of a single spin in a quantum dot. *Phys. Rev. B*. **69**, 1–8 (2004).

24. R. Jozsa, Fidelity for Mixed Quantum States. *J. Mod. Opt.* **41**, 2315–2323 (1994).

25. P. O. Boykin, T. Mor, M. Pulver, V. Roychowdhury, F. Vatan, New universal and fault-tolerant quantum basis. *Inf. Process. Lett.* **75**, 101–107 (2000).

26. M. Taherkhani, M. Willatzen, E. V. Denning, I. E. Protsenko, N. Gregersen, High-fidelity optical quantum gates based on type-II double quantum dots in a nanowire. *Phys. Rev. B: Condens. Matter Mater. Phys.* **99** (2019), doi:10.1103/PhysRevB.99.165305.

27. S. Nadj-Perge, S. M. Frolov, E. P. A. M. A. M. Bakkers, L. P. Kouwenhoven, Spin-orbit qubit in a semiconductor nanowire. *Nature*. **468**, 1084–1087 (2010).

28. E. V. Denning, J. Iles-Smith, N. Gregersen, J. Mork, Phonon effects in quantum dot single-photon sources. *Opt. Mater. Express*. **10**, 222 (2020).

29. J. Förstner, C. Weber, J. Danckwerts, A. Knorr, Phonon-assisted damping of rabi oscillations in semiconductor quantum dots. *Phys. Rev. Lett.* **91**, 1–4 (2003).

30. A. J. Ramsay, T. M. Godden, S. J. Boyle, E. M. Gauger, A. Nazir, B. W. Lovett, A. M. Fox, M. S. Skolnick, Phonon-induced Rabi-frequency renormalization of optically driven single InGaAs/GaAs quantum dots. *Phys. Rev. Lett.* **105**, 1–4 (2010).

31. A. Artioli, S. Kotal, N. Gregersen, P. Verlot, J. M. Gérard, J. Claudon, Design of Quantum Dot-Nanowire Single-Photon Sources that are Immune to Thermomechanical Decoherence. *Phys. Rev. Lett.* **123**, 247403 (2019).

32. L. Hackl, D. Li, N. Akopian, M. Christandl, Experimental proposal to probe the extended Pauli principle.

33. N. V. Vitanov, A. A. Rangelov, B. W. Shore, K. Bergmann, Stimulated Raman adiabatic passage in physics, chemistry, and beyond. *Rev. Mod. Phys.* **89**, 1–72 (2017).

34. A. M. Fox, Department of Physics and Astronomy Mark Fox, *Optical Properties of Solids* (Oxford University Press, 2001).

35. I. Vurgaftman, J. R. Meyer, L. R. Ram-Mohan, Band parameters for III-V compound semiconductors and their alloys. *J. Appl. Phys.* **89**, 5815–5875 (2001).

36. A. Reigue, J. Iles-Smith, F. Lux, L. Monniello, M. Bernard, F. Margaillan, A. Lemaitre, A. Martinez, D. P. S. S. McCutcheon, J. Mørk, R. Hostein, V. Voliotis, Probing Electron-Phonon Interaction through Two-Photon Interference in Resonantly Driven Semiconductor Quantum Dots. *Phys. Rev. Lett.* **118**, 1–6 (2017).

37. U. Bockelmann, G. Bastard, Phonon scattering and energy relaxation in two-, one-, and zero-dimensional electron gases. *Phys. Rev. B Condens. Matter*. **42**, 8947–8951 (1990).

38. P. Tighineanu, C. L. Dreeßen, C. Flindt, P. Lodahl, A. S. Sørensen, Phonon Decoherence of Quantum Dots in Photonic Structures: Broadening of the Zero-Phonon Line and the Role of Dimensionality. *Phys. Rev. Lett.* **120**, 257401 (2018).

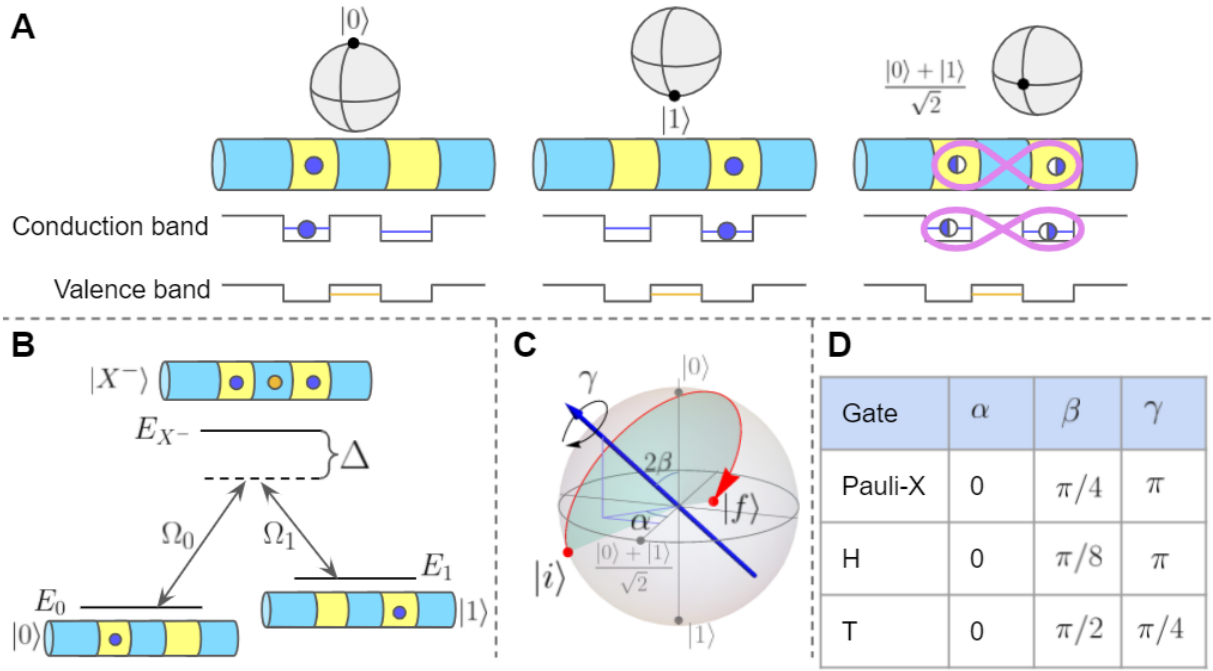

**Figure 1. The concept of a location qubit and the scheme for single-qubit manipulations. (A)** A location qubit is defined on a pair of neighbouring crystal-phase quantum dots. An electron(blue dot) can either be localized in the left or the right dot, constituting the $|0\rangle$ and $|1\rangle$ qubit states. An even superposition of the qubit states is shown on the right. For an InP nanowire, yellow(blue) regions represent ZB(WZ) crystal structure. **(B)** The qubit manipulation scheme shown as the energy level diagram of the $\Lambda$-system. The states $|0\rangle$ and $|1\rangle$ share a common spatially separated excited state $|X^-\rangle$, the negatively charged exciton state (where the orange dot is a hole that can form an exciton with either of the electrons on the left or right). Two laser pulses with Rabi frequency $\Omega_0$ and $\Omega_1$ drive the transitions respectively, both detuned for the same amount $\Delta$ from the common excited state. **(C)** Single-qubit manipulation. On the Bloch sphere, an arbitrary single-qubit gate corresponds to a clockwise rotation (red arrow) of the qubit state from initial state $|i\rangle$ to a final state $|f\rangle$ with respect to the rotation axis (blue arrow). The rotation axis is uniquely defined by $\alpha$ and $\beta$. Together with the angle of rotation $\gamma$ they determine which single-qubit gate is being implemented. **(D)** Representative single-qubit gates and the corresponding angles $\alpha$, $\beta$ and $\gamma$.

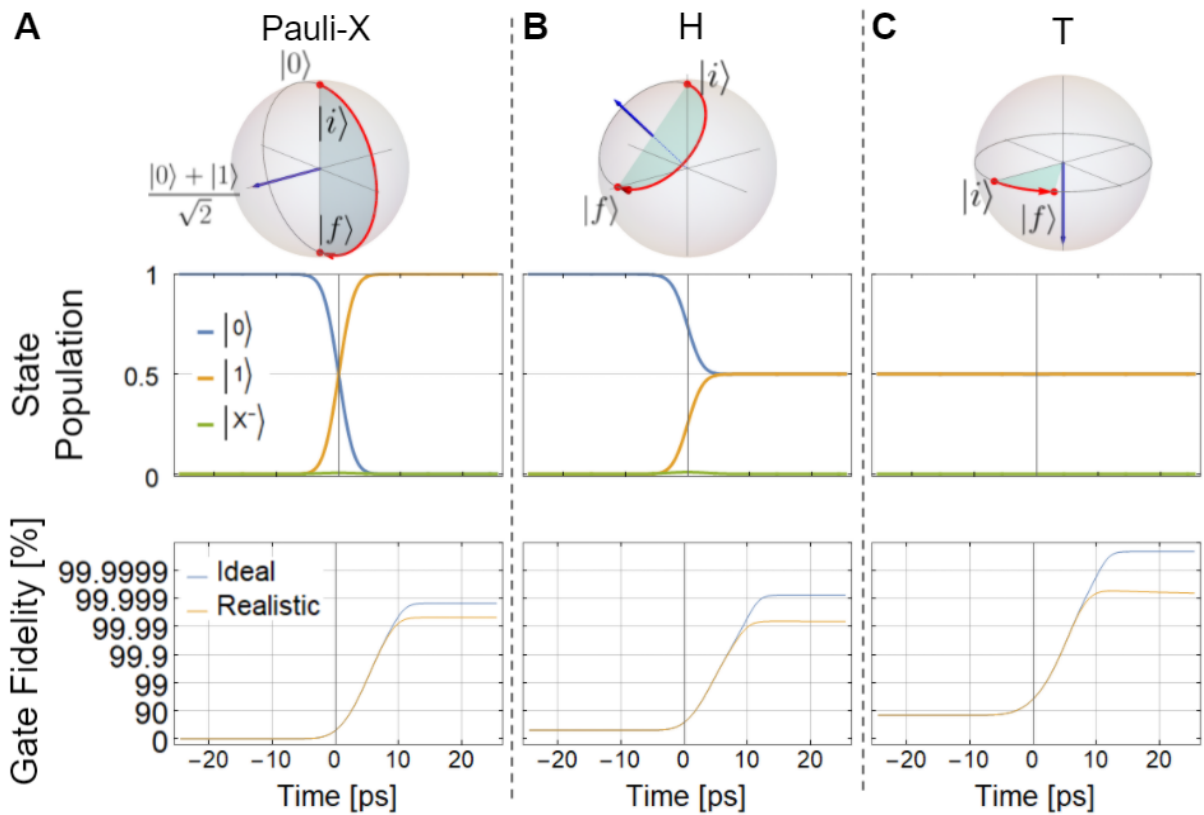

**Figure 2. State population and gate fidelity evolution for single-qubit gates.** For each representative single-qubit gate, on the Bloch spheres we have the rotation axis(blue arrow) and the rotation trajectory(red arrowed trace). The population of each state over the gating procedure is plotted in the middle row. In the gate fidelity plot, the ideal(realistic) fidelity is the simulated fidelity without(with) the effects of decoherence. **(A)** Pauli-X gate acting on the state $|0\rangle$, the resultant final state is the state $|1\rangle$, a full charge transfer has occurred. **(B)** Hadamard gate acting on the state $|0\rangle$, turning the state into an even superposition. **(C)** T gate acting on an even superposition of $|0\rangle$ and $|1\rangle$ states. Population is not affected since T merely adds a phase to the $|1\rangle$ state.

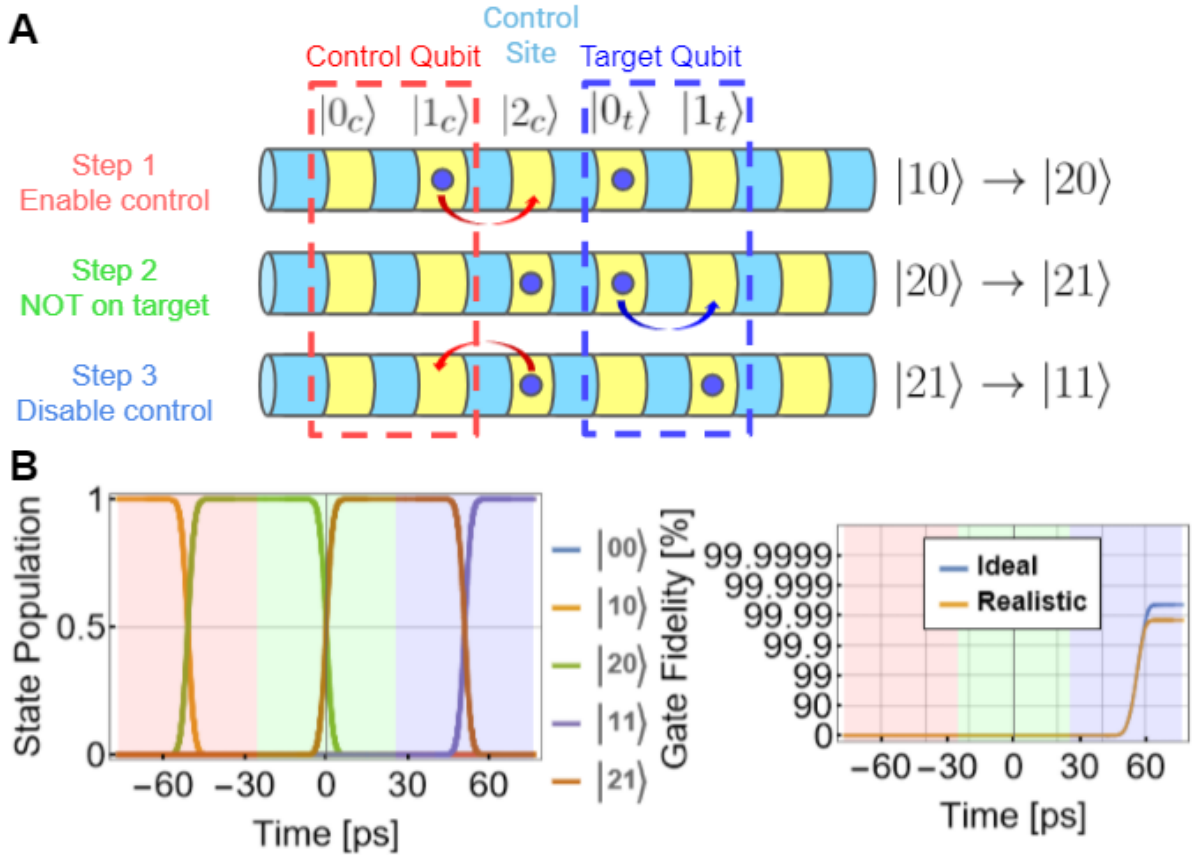

**Figure 3. The procedure of CNOT gate on location qubits. (A)** The CNOT gate exploits an additional quantum dot, acting as a control site between the control qubit and the target qubit. The procedure consists of three steps. In the first step, the control mechanism is activated by moving the electron on $|1_c\rangle$ to $|2_c\rangle$ (control site). With an electron on $|2_c\rangle$, a Coulomb potential exerted on the neighbouring quantum dots modifies the energy levels significantly on the nearest neighbours and very little on quantum dots further away (see main text), hence it differentiates the energy levels and transition energies on the target qubit with or without an electron initially in the $|1_c\rangle$ state. In the second step, a NOT(Pauli-X) gate is applied on the target qubit according to the modified energy levels. In the third step the control mechanism is disabled by moving the electron on the control site $|2_c\rangle$ back to $|1_c\rangle$. **(B)** The state population and **(C)** gate fidelity evolutions when CNOT gate acts on the $|1_c 0_t\rangle$ state.

Each shaded area corresponds to a step in the CNOT scheme of the same color as in **(A)**.

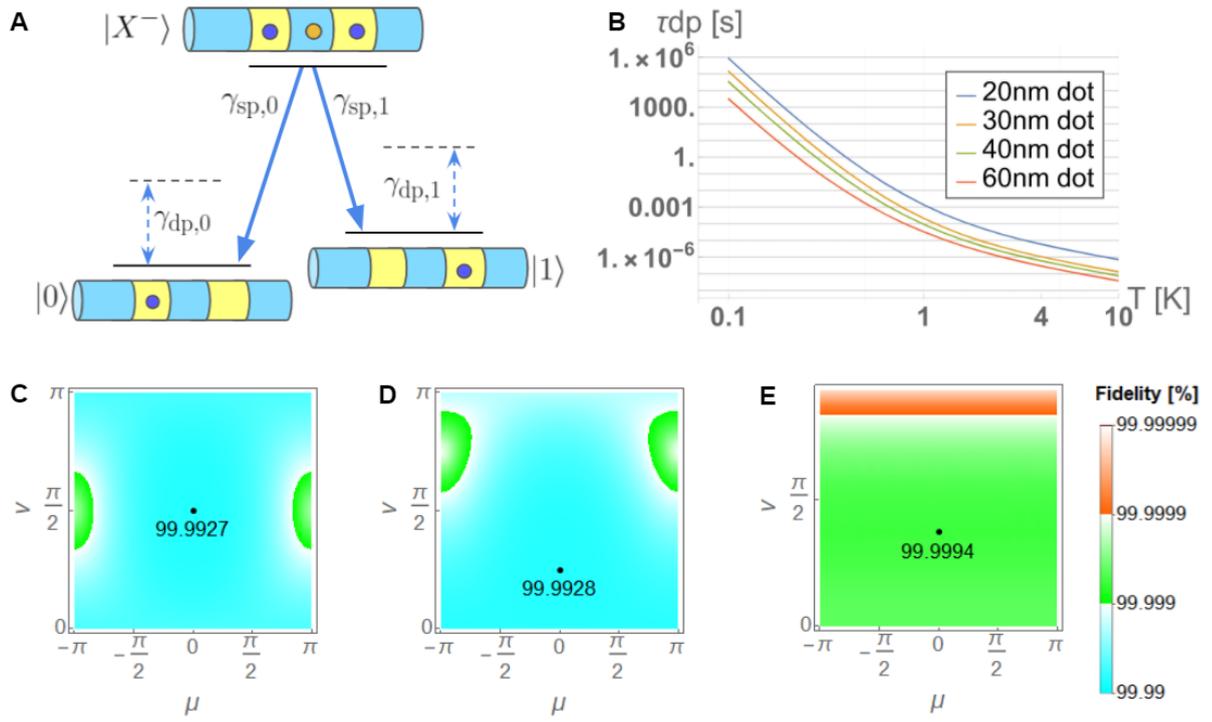

**Figure 4. Coherence time and realistic fidelities. (A)** The main decoherence mechanisms for the location qubits. The solid blue arrows represent spontaneous emission with respective rates $\gamma_{sp,0}$ and $\gamma_{sp,1}$. The dashed arrows represent dephasing due to virtual transitions induced by phonon scattering, characterized by the dephasing rates $\gamma_{dp,0}$ and $\gamma_{dp,1}$. **(B)** Dephasing time $\tau_{dp} = 1/\gamma_{dp}$ as a function of the operational temperature plotted for quantum dots of 4 different sizes. The nanowire used in the calculation is InP nanowire with a diameter of 20 nm. For a dot of 20 nm at 4K, the dephasing time is $6\mu s$. Fidelities of the **(C)** Pauli-X gate, **(D)** Hadamard gate and **(E)** T gate acting on an arbitrary initial state (parametrized as described in the main text). The lowest fidelities are indicated by the dots at 99.9927% ($X$), 99.9928% ($H$) and 99.9994% ($T$).

# Supplementary information - Location qubits in a multi-quantum-dot system

Dayang Li[1], Nika Akopian[1]*

[1] DTU Department of Photonics Engineering, Technical University of Denmark; Kongens Lyngby, 2800, Denmark

*Corresponding author. Email: nikaak@fotonik.dtu.dk

## A. The qubit manipulation scheme

The Hamiltonian of a generic $\Lambda$-type three level system takes the following form in the rotating wave approximation

$$H = \begin{bmatrix} 0 & \Omega_0(t)\exp(i\alpha) & 0 \\ \Omega_0(t)\exp(-i\alpha) & \Delta & \Omega_1(t) \\ 0 & \Omega_1(t) & 0 \end{bmatrix}.$$

The frequencies $\Omega_0$ and $\Omega_1$ are real, while the Rabi frequencies have a relative phase difference $\alpha$ that can be controlled by the constituent electric fields. The frequencies $\Omega_0$ and $\Omega_1$ are time dependent and have the following Gaussian temporal profiles with the same full width at half maximum of $2.355\sigma_\mathrm{P}$,

$$\Omega_0(t) = \Omega_{0,\max}\exp(-t^2/2\sigma_\mathrm{p}^2),$$

$$\Omega_1(t) = \Omega_{1,\max}\exp(-t^2/2\sigma_\mathrm{p}^2).$$

The normalized eigenstates and corresponding eigenenergies are given below.

$$\lambda_1 = 0$$
$$|\Phi_1(t)\rangle = -e^{i\alpha}\sin(\beta)|0\rangle + \cos(\beta)|1\rangle$$
$$\lambda_2 = -2Z(t)\sin^2(\phi(t))$$
$$|\Phi_2(t)\rangle = -e^{i\alpha}\cos(\beta)\cos(\phi)|0\rangle + \sin(\phi)|X^-\rangle - \sin(\beta)\cos(\phi)|1\rangle$$
$$\lambda_3 = 2Z(t)\cos^2(\phi(t))$$
$$|\Phi_3(t)\rangle = e^{i\alpha}\cos(\beta)\sin(\phi)|0\rangle + \cos(\phi)|X^-\rangle + \sin(\beta)\sin(\phi)|1\rangle$$

The following parameters are defined during the derivation.

$$\Omega_{\text{rms}}(t) = \sqrt{\Omega_0(t)^2 + \Omega_1(t)^2},$$

$$\tan(\beta) = \frac{\Omega_1(t)}{\Omega_0(t)},$$

$$Z(t) = \sqrt{\Omega_{\text{rms}}(t)^2 + (\Delta/2)^2},$$

$$\tan(2\phi(t)) = \frac{2\Omega_{\text{rms}}(t)}{\Delta}.$$

The definition of the angle $\phi(t)$ also implies the following relations:

$$\sin(2\phi(t)) = \Omega_{\text{rms}}(t)/\sqrt{\Omega_{\text{rms}}(t)^2 + (\Delta/2)^2} = \Omega_{\text{rms}}(t)/Z(t),$$

$$\cos(2\phi(t)) = 1 - 2\sin^2(\phi(t)) = (\Delta/2)/\sqrt{\Omega_{\text{rms}}(t)^2 + (\Delta/2)^2} = (\Delta/2)/Z(t),$$

$$2\sin^2(\phi(t)) = 1 - (\Delta/2)/Z(t).$$

As discussed in the main text, the common excited state $|X^-\rangle$ is the one subject to most decoherence processes. In order to avoid populating this state, we notice that the first eigenstate $|\Phi_1(t)\rangle$ contains no $|X^-\rangle$ component, and the second eigenstate $|\Phi_2(t)\rangle$ contains a small $|X^-\rangle$ component so long as $\phi$ is small. The condition that $\phi$ is small indicates $\Omega_{\text{rms}} \ll \Delta$, since $\tan(2\phi) \approx 2\phi = 2\Omega_{\text{rms}}/\Delta$ for $\phi \ll 1$. Since the root mean square Rabi frequency $\Omega_{\text{rms}}$ is time dependent, the condition $\Omega_{\text{rms}} \ll \Delta$ should be held at all times, especially for its maximum $\Omega_{\text{rms,max}} = \sqrt{\Omega_{0,\text{max}}^2 + \Omega_{1,\text{max}}^2}$. The condition $\phi \ll 1$ will be taken as an assumption in the following derivation. This also leads to $\cos(\phi) \approx 1$ and $\sin(\phi) \approx 0$. Under the assumption, the eigenstates thus become the following.

$$|\Phi_1(t)\rangle = \sin(\beta)|0\rangle - e^{-i\alpha}\cos(\beta)|1\rangle$$
$$|\Phi_2(t)\rangle = \cos(\beta)|0\rangle + e^{-i\alpha}\sin(\beta)|1\rangle$$
$$|\Phi_3(t)\rangle = |X^-\rangle$$

Since the excited state $|X^-\rangle$ can spontaneously emit to the two lower lying states, the excited state component should generally be avoided. Two of the eigenstates $|\Phi_1(t)\rangle$ and $|\Phi_2(t)\rangle$ have almost no excited state component (dark states). Therefore by decomposing any qubit state as a linear combination of the dark states, decoherence can be significantly minimized at all stages of operation, especially during gating procedures. When one of the dark state components adiabatically obtains a dynamic phase relative to the other, a single qubit gate is achieved. On the Bloch sphere, it is represented by a rotation. (main text Fig. 1(c)) An arbitrary initial qubit state $|\psi\rangle$ at time $t = t_0$ can be expressed as a linear combination of the eigenstates $|\psi\rangle = \langle\Phi_1|\psi\rangle|\Phi_1\rangle + \langle\Phi_2|\psi\rangle|\Phi_2\rangle$. At a later time $t$, the qubit state evolves to $|\psi(t)\rangle = \langle\Phi_1|\psi\rangle|\Phi_1\rangle + e^{-i\int_{t_0}^{t}\lambda_2(t')dt'}\langle\Phi_2|\psi\rangle|\Phi_2\rangle$. A phase factor $e^{-i\int_{t_0}^{t}\lambda_2(t')dt'}$ is gained by the $|\Phi_2\rangle$ component relative to the $|\Phi_1\rangle$ component. The phase factor $e^{-i\int_{t_0}^{t}\lambda_2(t')dt'}$ depends solely on the single-photon detuning $\Delta$, if the Rabi frequency envelopes are kept unchanged, and the time $t$ is after the driving field has been switched off. Hence by varying $\Delta$ one can control the phase factor and implement specific single qubit rotations. An intuitive description on the Bloch sphere can be given. The state $|\Phi_2\rangle = \cos(\beta)|0\rangle + e^{-i\alpha}\sin(\beta)|1\rangle$ is located on the Bloch sphere with the following coordinates (Bloch vector), $(\sin(2\beta)\cos(-\alpha), \sin(2\beta)\sin(-\alpha), \cos(2\beta))$. The orthogonal state $|\Phi_1\rangle = \sin(\beta)|0\rangle - e^{-i\alpha}\cos(\beta)|1\rangle$ is located directly on the opposite end of the Bloch sphere. These two states define the rotation axis around which the qubit state will rotate. To be more specific, the rotation axis is parallel with $|\Phi_2\rangle$ (antiparallel with $|\Phi_1\rangle$). An initial qubit state $|\psi\rangle$ is first projected onto the two eigenstates. Then a phase is gained by the $|\Phi_2\rangle$ component relative to $|\Phi_1\rangle$. On the bloch sphere, this is a clockwise rotation of the qubit state relative to $|\Phi_2\rangle$ by an angle $\gamma$. The full expression of rotation angle $\gamma$ is given by

$$\gamma = -\int_{t_0}^{t} \lambda_2(t')\mathrm{d}t' = \int_{t_0}^{t} 2\sin^2(\phi(t'))Z(t')\mathrm{d}t'$$
$$= \int_{t_0}^{t} (\sqrt{\Omega_{rms}^2(t') + (\Delta/2)^2} - \Delta/2)\mathrm{d}t'$$

Once the qubit rotation axis is determined, one will know the required temporal envelopes of the Rabi frequencies. So the rotation angle $\gamma$ is uniquely determined by the single photon detuning $\Delta$. The rotation angle with respect to the rotation axis, $\gamma(\Delta)$, depends on the single photon detuning $\Delta$ as shown in Figure S1 However in an experimental setup it is often the other way around. We choose the single photon detuning $\Delta$ according to the angle of rotation that needs to be implemented.

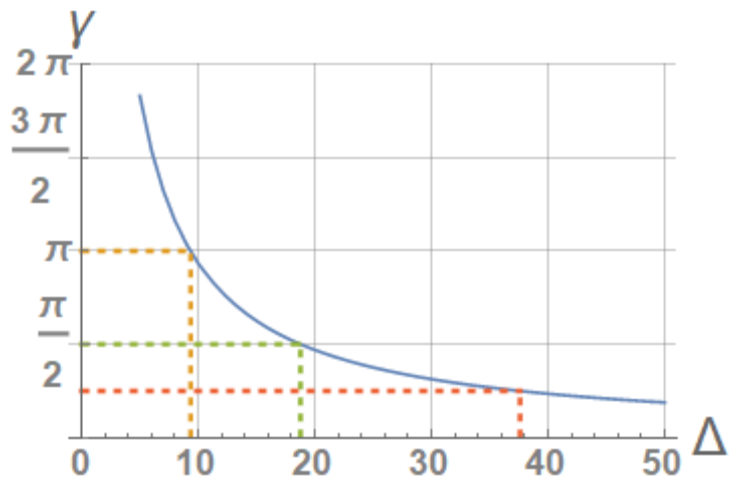

Figure S1 The rotation angle $\gamma$ as a function of the single photon detuning $\Delta$ in units of the maximal root mean square Rabi frequency $\Omega_{\rm rms,max}$.

In addition, throughout the single qubit rotation process, the adiabatic condition shall be satisfied. The qubit control scheme (STIRSAP) follows the same adiabatic constraints(*33*) as the stimulated Raman adiabatic passage(STIRAP).

$$\Omega_{\rm rms,max}\tau_{\rm p} \geq \mathcal{A}_{\rm min}$$

In the numerical simulations of this work, a pulse duration of $\tau_{\rm p} = 6\sigma_{\rm p}$ is used. The minimal pulse area $\mathcal{A}_{\rm min}$ is chosen to be 100 for high gate performance, it can be as low as 10 in principle(*33*). Thus the relevant frequency scales of the system should satisfy the following.

$$\Omega_{\rm rms,max}\tau_{\rm p} \geq \mathcal{A}_{\rm min} = 100 \gg \pi/2, \qquad \Omega_{\rm rms}(t) \ll \Delta.$$

In essence, the STIRSAP technique provides a scheme to arbitrarily rotate a single qubit state. The rotation axis is defined by the two eigenstates $|\Phi_1\rangle$ and $|\Phi_2\rangle$ of the STIRSAP Hamiltonian, which further depends on the two angles $\alpha$ and $\beta$. The $\alpha$ angle is the phase difference of the two Rabi frequencies, which originates from the phase difference between the two electric fields. The angle $\beta$ is defined in terms of the ratio between the two Rabi frequencies, $\tan(\beta) = \Omega_1/\Omega_0$. Since the Rabi frequencies are proportional to the constituent electric field strengths, they are also proportional to the square root of the driving laser intensities $I_0(t)$ and $I_1(t)$ that are experimentally tunable, such that $\tan(\beta) = \sqrt{I_1(t)/I_0(t)}$.

## B. Decoherence analysis

The decoherence mechanisms following Markovian dynamics are treated with the following Lindblad master equation

$$\partial_t \rho_S = -i[H_S, \rho_S] + \sum_{j=0,1} D(\Gamma_{\text{sp},j}, L_{X^- \to j})\rho_S + \sum_{j=0,1} D(\Gamma_{\text{dp}}, L_{\text{dp},j})\rho_S.$$

The system is described by the reduced density operator $\rho_S$. $D(\Gamma, L)$ is the Lindblad superoperator given as $D(\Gamma, L)\rho_S = \Gamma \left( L\rho_S L^\dagger - \frac{1}{2}(L^\dagger L \rho_S + \rho_S L^\dagger L) \right)$, where $\Gamma$ is the rate governing the decoherence process described by $L$. The term describing spontaneous emission from $|X^-\rangle$ to $|0\rangle (|1\rangle)$ is specified by the spontaneous emission rates $\Gamma_{\text{sp},X^- \to 0}(\Gamma_{\text{sp},X^- \to 1})$ and the following process operators (fermionic annihilation operators in the given basis)

$$L_{X^- \to 0} = \begin{bmatrix} 0 & 1 & 0 \\ 0 & 0 & 0 \\ 0 & 0 & 0 \end{bmatrix} \left( L_{X^- \to 1} = \begin{bmatrix} 0 & 0 & 0 \\ 0 & 0 & 0 \\ 0 & 1 & 0 \end{bmatrix} \right).$$

The phononic dephasing term is specified by the dephasing rate $\Gamma_{dp}$ and the following process operators

$$L_{\text{dp},0} = \begin{bmatrix} 1 & 0 & 0 \\ 0 & 0 & 0 \\ 0 & 0 & 0 \end{bmatrix} \left( L_{\text{dp},1} = \begin{bmatrix} 0 & 0 & 0 \\ 0 & 0 & 0 \\ 0 & 0 & 1 \end{bmatrix} \right).$$

For larger systems, such as a two qubit system, the computation basis is expanded to the following $\{|0\rangle, |X_{01}^-\rangle, |1\rangle, |X_{12}^-\rangle, |2\rangle\} \otimes \{|0\rangle, |X_{01}^-\rangle, |1\rangle\}$. The Lindblad operators should be modified accordingly. For instance, dephasing of state $|1\rangle$ of the first qubit should act on all states with the first qubit being $|1\rangle$. The Lindblad operator is a diagonal matrix of the

form $L_{\mathrm{dp},11} = \mathrm{Diag}(0,0,1,0,0,0,0,1,0,0,0,0,1,0,0)$. The same principle applies to the Lindblad operators characterizing spontaneous emission.

In the following analysis, the material platform is chosen as an InP nanowire with alternating WZ and ZB crystal phases. Both the conduction band and the valence band in the WZ crystal phase have a higher energy than the ZB phase, therefore the electrons are trapped in the ZB sections and the holes in WZ sections.

## B.1 The spontaneous emission rate

The spontaneous emission rate

$$\Gamma_{\mathrm{sp}} = \frac{d^2 \omega^3}{3\pi \epsilon_0 \hbar c^3}$$

is evaluated for realistic device dimensions (ensuring electron localization). Where $d$ is the dipole matrix element between the excited and the ground state, $\omega$ is the transition energy, $\epsilon_0$ the vacuum permittivity and $c$ the speed of light. The spontaneous emission rates are calculated by first finding the wave functions of the charge carriers and the corresponding energies, then the dipole matrix elements are found, lastly everything is put together.

The treatment for charge carrier wavefunctions in quantum confinement structures follows the general single-particle approach as in Fox(*34*), similar to the treatment for self-assembled quantum dots. The relevant material parameters can be found in Table S1.

| Parameter | Symbol | Value |
|---|---|---|
| Effective mass for electron | $m_e$ | $0.067 m_0$ |
| Effective mass for hole | $m_h$ | $0.64 m_0$ |
| Bandgap ZB | $E_{g,ZB}$ | 1.410eV |
| Bandgap WZ | $E_{g,WZ}$ | 1.474eV |
| Conduction band offset | $\Delta_c$ | 129meV |
| Valence band offset | $\Delta_v$ | 65meV |
| Momentum matrix element | $M^2$ | 10.35eV·$m_0$(35) |
| Refractive index near the bandgap | $n$ | 3.44(4) |

Table S1 Material parameter for InP.

The system under consideration is a pair of quantum dots as shown in Figure S2. For an electron, the device geometry is a rotationally symmetric cylindrical WZ nanowire with two ZB sections inserted.

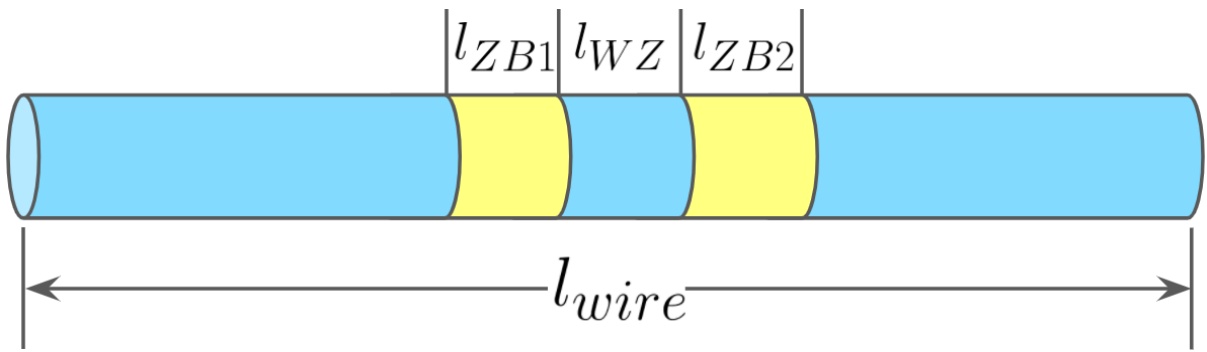

Figure S2: Device geometry for a pair of neighbouring quantum dots for electrons. The blue regions are in the WZ crystal phase while the red sections are in ZB.

The time independent Schrodinger equation

$$\left(-\frac{\hbar^2}{2m_{e(h)}} + V_{e(h)}(\mathbf{r})\right)\Psi_{e(h)}(\mathbf{r}) = E\Psi_{e(h)}(\mathbf{r})$$

is solved in this geometry, with the boundary condition that the wavefunctions vanish at the surface of the wire. The radial confinement potential is assumed to be a cylindrically symmetric infinite potential well, while the confinement potential for electrons(holes) along the wire is taken as finite potential wells formed by conduction(valence) band alignment. Due to the separability of the wavefunctions, a solution can be decomposed as $\Psi_{e(h)}(\mathrm{r}) = \Theta(\theta)R(r)Z(z)$. One can obtain an equation for the radial, azimuthal and axial wavefunctions respectively.

$$\frac{\mathrm{d}^2}{\mathrm{d}\theta^2}\Theta(\theta) = -m^2\Theta(\theta),$$

$$\left(\frac{\mathrm{d}^2}{\mathrm{d}r^2} + \frac{1}{r}\frac{\mathrm{d}}{\mathrm{d}r} + \frac{m^2}{r^2}\right)R(r) = -\frac{2m_e}{\hbar^2}(E_r - V_r)R(r),$$

$$\frac{\mathrm{d}^2}{\mathrm{d}z^2}Z(z) = -\frac{2m_e}{\hbar^2}(E_z - V_{z,e}(z))Z(z).$$

The energy eigenvalue of $\Psi_{e(h)}$ is given as $E = E_r + E_z$.

**The radial wavefunctions**

The normalized radial wavefunctions are proportional to the Bessel function of the first kind $J_m(\kappa r)$ with $\kappa = \sqrt{2m_e E_r}/\hbar$ for $r$ smaller than the radius of the wire $r_w$. The corresponding eigenenergies are given by $E_r^{ml}$.

$$R^m(r) = \frac{J_m(\kappa r)}{\sqrt{\int_0^{r_w} J_m^2(\kappa r) r \, \mathrm{d}r}}$$

$$E_r^{ml} = \frac{\hbar^2}{2 m_e r_w^2} j_{m,l}$$

$j_{m,l}$ denotes the $l$'th zero of $J_m(\kappa r)$. The ground state has energy 13.37meV assuming a wire radius of 10nm.

**The angular wavefunctions**

The normalized angular wavefunctions are of the form $\Theta(\theta) = e^{im\theta}/\sqrt{2\pi}$ with the corresponding eigenvalues $-m^2$, where $m \in \{-l, -l+1, ..., l\}$. For the groundstate wavefunction, $l$ takes the value 0, therefore the angular wavefunction is $\Theta(\theta) = 1/\sqrt{2\pi}$.

**The axial wavefunctions**

The axial wavefunctions can be solved numerically in the given geometry by using finite element method on the device geometry. As an example, for a structure with $\{l_{ZB1}, l_{WZ}, l_{ZB2}\} = \{10\text{nm}, 50\text{nm}, 9\text{nm}\}$, the electron and hole wavefunctions are found and shown in Figure S3. The hole wavefunctions are found with the same approach by switching out the effective mass and band offsets accordingly.

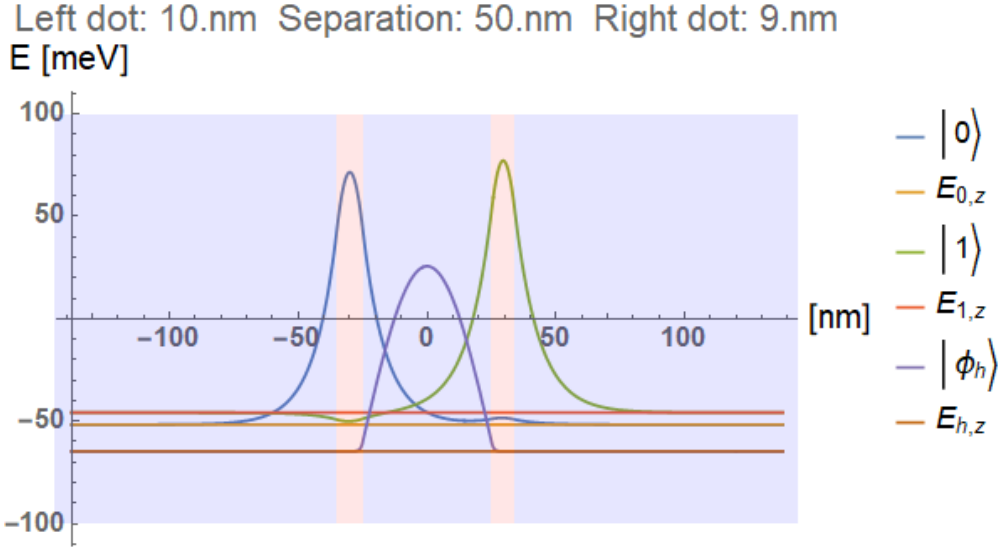

Figure S3 The axial wavefunction for a sample structure of dimensions {10nm, 50nm, 9nm}. As we can see, the wavefunctions still extend slightly into the other quantum dot, localization is not optimal, an undesirable finite interdot coupling arises.

The energy of an electron $E_e$ is measured from the conduction band for WZ, while the energy of a hole $E_h$ is measured from the valence band for WZ. The transition energies are given by $\hbar\omega = E_{g,WZ} + E_e + E_h$.

The dipole matrix element is ready to be evaluated. The dipole matrix element(5) between an electron and a hole state $|e\rangle$ and $|h\rangle$ is defined as $\mathbf{d} = \langle e|(-e)\mathbf{r}|h\rangle$. It is related to the momentum matrix element $M = \langle e|\mathbf{p}|h\rangle$ via $d^2 = e^2 M^2/(m_0^2 \omega^2)\langle Z_e|Z_h\rangle^2$. The momentum matrix element for many materials are measured experimentally and tabulated. For InP, the tabulated value(35) is $M^2/m_0 = 10.35 eV$.

By the analysis above, the dipole moment and hence the spontaneous emission rate can be calculated. For the transition involving the electron state $|0\rangle$ and the hole state $|h\rangle$, the spontaneous emission rate is found to be $2.57 \cdot 10^8 \text{s}^{-1}$, while for the transition involving $|1\rangle$ and $|h\rangle$, the spontaneous emission rate is $2.74 \cdot 10^8 \text{s}^{-1}$. The spontaneous emission rate is calculated for various structure dimensions $l_{ZB1}$, $l_{WZ}$, $l_{ZB2}$. $l_{ZB2}$ is fixed at 19nm, $l_{ZB1}$ varies between 20nm and 50nm, $l_{WZ}$ varies between 50nm and 200nm. The spontaneous emission rate is found within the interval $[9.27 \cdot 10^7 s^{-1}, 7.79 \cdot 10^8 s^{-1}]$. A larger quantum dot yields a smaller dipole moment due to a smaller overlap between the electron and hole wave functions, hence a smaller spontaneous emission rate. So in principle a larger quantum dot can help against spontaneous emission.

## B.2 Electron-phonon interaction

The coupling of a quantum dot to a phonon bath can be described by the Hamiltonian $H_I = H_L + H_Q$, where $H_L$ and $H_Q$ are the linear and quadratic coupling terms.

$$H_L = \hbar \sum_{\mathbf{k}} \left( g_{\mathbf{k}} |0\rangle\langle 1| b_{\mathbf{k}}^\dagger + H.c. \right)$$

$$H_Q = \hbar \sum_{\mathbf{k},\mathbf{k}',i=0,1} f_{\mathbf{k},\mathbf{k}'} |i\rangle\langle i| (b_{\mathbf{k}}^\dagger + b_{\mathbf{k}})(b_{\mathbf{k}'}^\dagger + b_{\mathbf{k}'})$$

As mentioned in the main text, the coupling mechanism is deformation coupling to bulk 3D longitudinal acoustic phonons.

The linear and quadratic coupling constants take the following forms(6).

$$g_{\mathbf{k}} = \langle 1|M_{\mathbf{k}}\varrho(\mathbf{k})|0\rangle$$

$$f(\mathbf{k}, \mathbf{k}', i) = \sum_{m>1} \frac{\langle i|M_{\mathbf{k}}\varrho(\mathbf{k})|m\rangle\langle m|M_{\mathbf{k}'}\varrho(\mathbf{k}')|i\rangle}{\omega_i - \omega_m}$$

where

$$M_{\mathbf{k}} = kD_e/\sqrt{2\rho\mathcal{V}\nu_{\mathbf{k}}}$$

is the phonon matrix element, and

$$\varrho(\mathbf{k}) = \int \mathrm{d}\mathbf{r} e^{i\mathbf{k}\cdot\mathbf{r}}\rho(\mathbf{r})$$

is the charge density operator in the $\mathbf{k}$-space where

$$\rho(\mathbf{r}) = \sum_{j,j'} \psi_j*(\mathbf{r})\psi_{j'}(\mathbf{r})c_j^\dagger c_{j'}$$

is the real space charge density. $D_e$ is the deformation coupling constant for electrons in ZB InP, $\rho$ is the material density, $c_s$ the speed of sound, $\mathcal{V}$ the quantization volume. By substituting the above definitions into the coupling constants, we obtain the following.

$$g_{\mathbf{k}} = \sqrt{\frac{\hbar k}{2\rho c_s \mathcal{V}}} D_e \langle 1|e^{i\mathbf{k}\cdot\mathbf{r}}|0\rangle$$

Here $|0\rangle$ and $|1\rangle$ refers to the qubit states. This is to be distinguished with the state references in the quadratic coupling constants.

The quadratic coupling between an electronic state and the phonon bath characterizes the following process. A phonon would initially be absorbed by an electron, make an attempt to drive an electronic transition to an excited state. However for crystal phase QDs, the energy difference between the one-electron ground state and the first excited state is on the order of tens of meVs, while the energy of thermal phonons at cryogenic temperatures are typically on the order of hundreds of $\mu\text{eVs}$. Due to the difference of energy scales, electronic transitions would not happen. Instead it leads to virtual transitions during which an electron is scattered into another mode of the same energy, while the electron in the one-electron ground state receives a random phase kick. The quadratic coupling constant is given by the following expression(*36*).

$$f(\mathbf{k},\mathbf{k}',i) = \sum_{m\geq 1} \frac{\langle 0|M_{\mathbf{k}}\varrho(\mathbf{k})|m\rangle\langle m|M_{\mathbf{k}'}\varrho(\mathbf{k}')|0\rangle}{\omega_0 - \omega_m}$$

It should be noted that the state $|0\rangle$ refers to the one-electron ground state on a single QD and $|m\rangle$ refers to excited states on the same QD.

**The linear coupling term**

The linear term describes an electronic transition accompanied by the emission or absorption of a phonon. This process is either 1) spontaneous and continuously affecting the qubit or 2) passive and only occurs during gating operation.

The spontaneous emission of phonons can be characterized by the spontaneous phonon emission rate. The phonon absorption/emission rate is given(*37*) based on Fermi's golden rule.

$$\gamma_{\text{phonon,abs/emis}} = \frac{2\pi}{\hbar} \sum_{f,\mathbf{k}} \alpha(\mathbf{k})^2 |\langle \psi_f | e^{\mp i\mathbf{k}\cdot\mathbf{r}} | \psi_i \rangle|^2 \delta(E_f - E_i \pm E_k) \left[ n_B(E_k,T) + \left\{ \begin{array}{c} 1 \\ 0 \end{array} \right\} \right]$$

In the expression, the upper(lower) sign is for phonon absorption(emission). The curly bracket term takes the value 1(0) for absorption(emission). Deformation potential coupling with LA phonons is included in the wavevector dependent parameter $\alpha(\mathbf{k})^2 = D_e^2 \hbar k/(2\rho c_s \mathcal{V})$. $n_B(E_k,T)$ is the bosonic occupation number for a phonon with energy $E_k$ at temperature $T$. Since at cryogenic temperatures ($T=4\mathrm{K}$) the occupation number of phonons at any energy is significantly smaller than unity, the spontaneous phonon emission process dominates over absorption events. The initial state under consideration is $|\psi_i\rangle=|1\rangle$ and the only final state is $|\psi_f\rangle=|0\rangle$. It is apparent that the phonon emission process is accompanied by an electronic transition. By using the wavefunctions calculated for crystal phase quantum dots, the spontaneous phonon emission rate is numerically calculated. The results as shown in Figure S4 indicates that a separation of 80-100nm is sufficient to eliminate this effect.

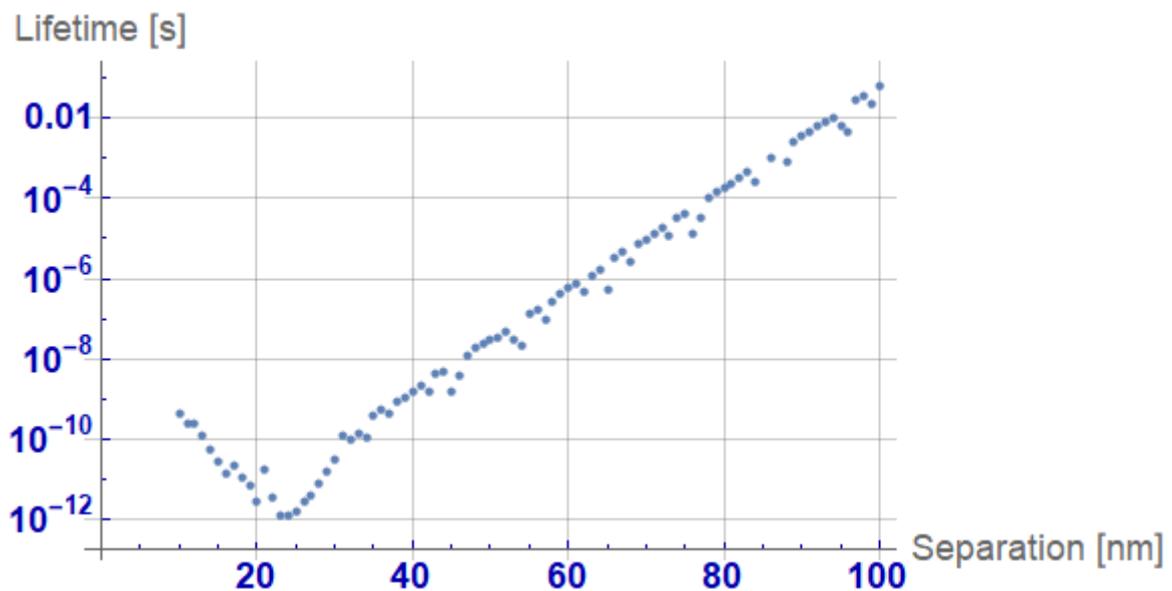

Figure S4 The phonon emission lifetime due to spontaneous emission of phonon. The calculations are based on two crystal phase QDs of 20 and 19nm respectively separated by a variable inter-dot distance (separation). The temperature is set at $T=4\mathrm{K}$. At a separation of 80nm or larger, the lifetime is long enough for millions of single qubit operations.

The passive process is a process describing the rearrangement of the crystal lattice in response to the modified charge density. This is described by the Franck-Condon principle, which states that the most probable transition between vibronic states are those that have a large initial and final phonon bath equilibrium overlap (Franck-Condon factor $B^2$ between $|0\rangle$ and $|1\rangle$). This means that the proportion $B^2$ is transferred without a phonon transition. The Franck-Condon factor of quantum dot states coupled to LA phonons via deformation potential coupling is found in (*28*) as

$$B = \exp\left(-\frac{1}{2}\int_0^\infty \nu^{-2} J_{PH}(\nu) \coth(\beta\hbar\nu/2)\mathrm{d}\nu\right),$$

where $\nu$ is the phonon energy,

$$J_{PH}(\nu) = \sum_{\mathbf{k}} |g_{\mathbf{k}}|^2 \delta(\nu - \nu_{\mathbf{k}})$$

is the phonon spectral density, and $\beta = (k_B T)^{-1}$.

Using the coupling constant derived earlier

$$g_{\mathbf{k}} = \sqrt{\frac{\hbar k}{2\rho c_s \mathcal{V}}} D_e \langle 1|e^{i\mathbf{k}\cdot\mathbf{r}}|0\rangle,$$

one can calculate the Franck-Condon factor for the transition between $|0\rangle$ and $|1\rangle$. Test calculations for two DQD dimensions (in nm) {20,50,19} and {10,100,9} over a range of temperatures are shown in Figure S5.

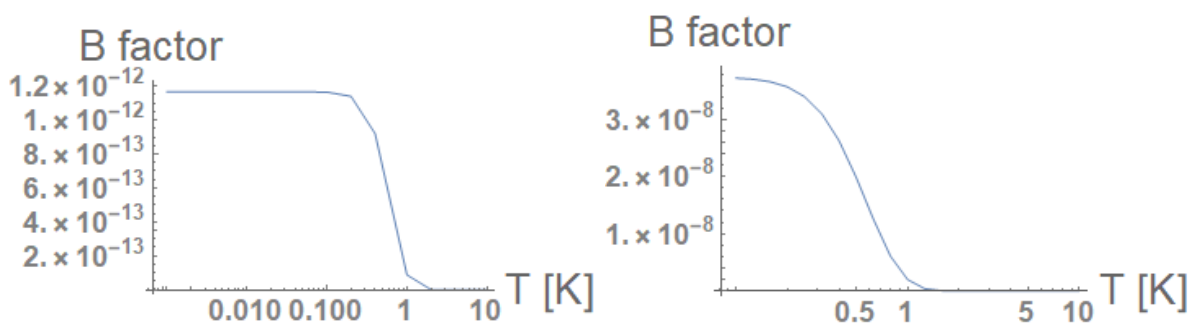

Figure S5 Test calculations of the B factor for two DQD dimensions (in nm) {20,50,19}(left) and {10,100,9}(right).

At temperatures lower than $T = 1K$, we notice that the B factors saturate on the order of $10^{-9}$ and $10^{-7}$. It means that the FC factors $B^2$ is negligibly small. Leading to the conclusion that almost the entire population that is being transferred during any single qubit gate would be incoherent due to the vast difference in the two equilibrium phonon configurations. If this were to be true, other solid state charge qubits that have spatially distinct wavefunctions would also suffer from this process. Here a discrepancy between theoretical predictions and actual observations persist. Experimental results from other solid state charge qubits do not exhibit limitations of this nature, even a single qubit gate fidelity of 86% is obtainable(*9*), and the dominant limitation there is claimed to be charge noise. If the conclusion drawn earlier would be true, no coherent charge transfer would be observable in solid state charge qubits at all. Based on the experimental observations, the authors predict that electronic transition accompanied by phonon transition can not be a detrimental factor to the scheme, yet a quantitative treatment requires finer models of the system. But this does not affect the main message of this manuscript.

**The quadratic coupling term**

The quadratic coupling to the phonon bath gives rise to virtual transitions. The physical picture is that a phonon will make an attempt to drive the charge states to higher energy states. However a phonon at a typical cryogenic temperature of 1K has an energy on the order of 0.1meV, while the energy difference between the qubit 0 and 1 state and the higher lying energy levels are typically 80-100meV. Due to the difference in energy scale, no real transition will occur, instead phonons will be absorbed and emitted again into a different mode of the same energy, inducing a random phase disruption to the electronic state. As a result, dephasing occurs on each of the charge states. This is characterized by the dephasing rate, obtainable from two independent methods (polaron transform(*36*) and cumulant expansion(*38*)).

$$\gamma_{\rm dp} = \frac{\pi V^2}{c_s(2\pi)^6}\int d\mathbf{k}d\mathbf{k'}|f(\mathbf{k},\mathbf{k'})|^2\{n(k')[n(k)+1]+n(k)[n(k')+1]\}\delta(k-k')$$

where $f(\mathbf{k},\mathbf{k'})$ is given by the following, if we only consider contributions to virtual transition to the lowest excited state.

$$f(\mathbf{k},\mathbf{k'},i) = \frac{\langle 0|M_{\mathbf{k}}\varrho(\mathbf{k})|1\rangle\langle 1|M_{\mathbf{k'}}\varrho(\mathbf{k'})|0\rangle}{\omega_0 - \omega_1}$$

The axial wavefunctions of the states $|0\rangle$ and $|1\rangle$ are given in Figure S6. By using the above wavefunctions, the dephasing rates are evaluated for various dot sizes over a range of temperatures. The results are presented in Figure 4(b) in the main text.

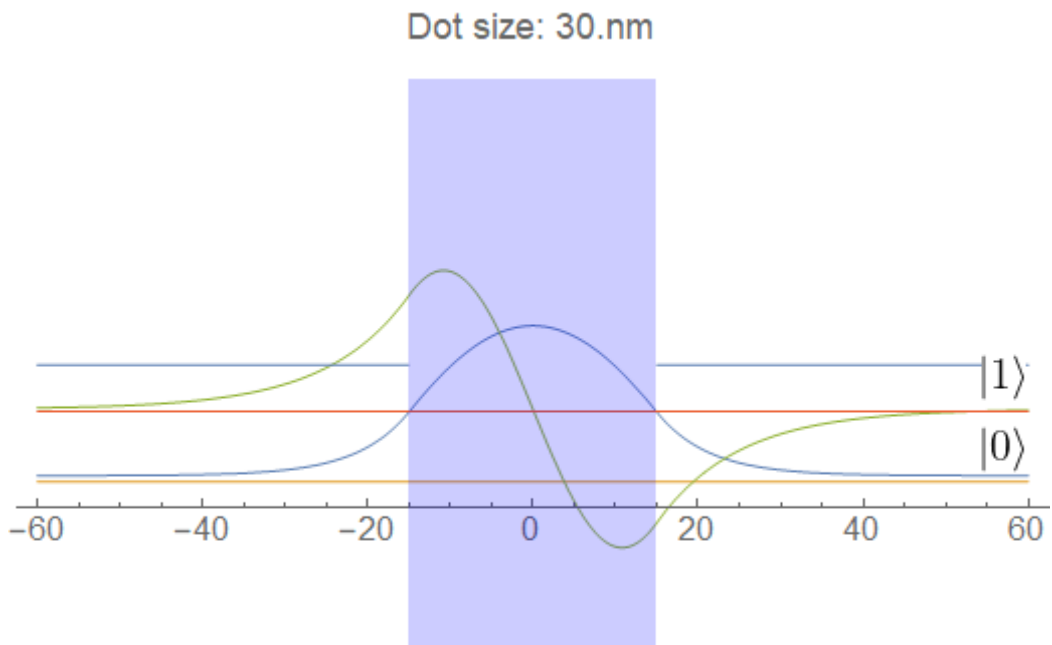

Figure S6 The axial wavefunctions of the ground state in a single crystal phase QD and the first excited state. A dot size of 30 nm is chosen for the illustration. The blue flat line

indicates the confinement potential, zero energy coincides with the conduction band edge of the ZB section.

## C. CNOT gate acting on more general two-qubit initial states

In this section, the CNOT gate is applied on a range of test initial states. The test states are representative as they contain complex linear combinations of the two-qubit states $\{|0_c0_t\rangle, |0_c1_t\rangle, |1_c0_t\rangle, |1_c1_t\rangle\}$. Since the CNOT gate consists of a series of single qubit gates, the gate fidelitys are generally lower than individual single qubit gates. In our simulations, the CNOT gate works also for the states that are supposedly preserved. The results are presented in Figures S7-S9.

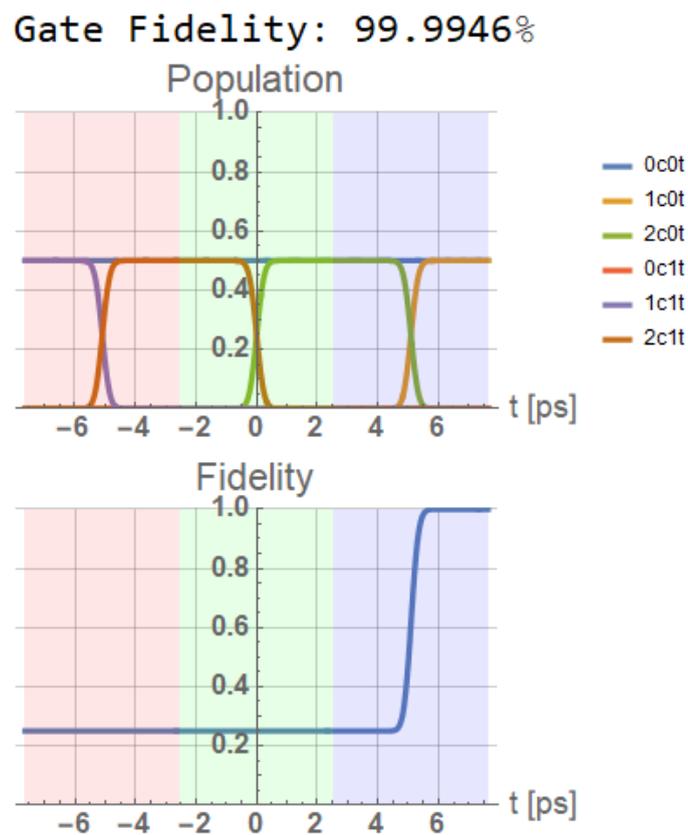

Figure S7 CNOT gate acting on $\frac{1}{\sqrt{2}}(|00\rangle + |11\rangle)$, the resultant state is $\frac{1}{\sqrt{2}}(|00\rangle + |10\rangle)$ as shown by the population and fidelity.

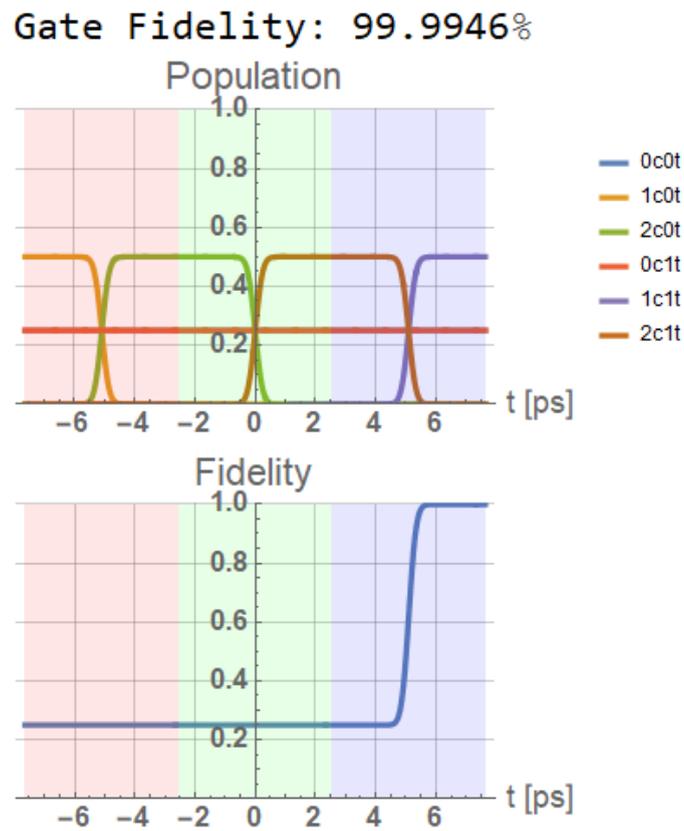

Figure S8 CNOT gate acting on $\frac{1}{2}(|00\rangle + |01\rangle) + \frac{1}{\sqrt{2}}|10\rangle$, the resultant state is $\frac{1}{2}(|00\rangle + |01\rangle) + \frac{1}{\sqrt{2}}|11\rangle$ as shown by the population and fidelity.

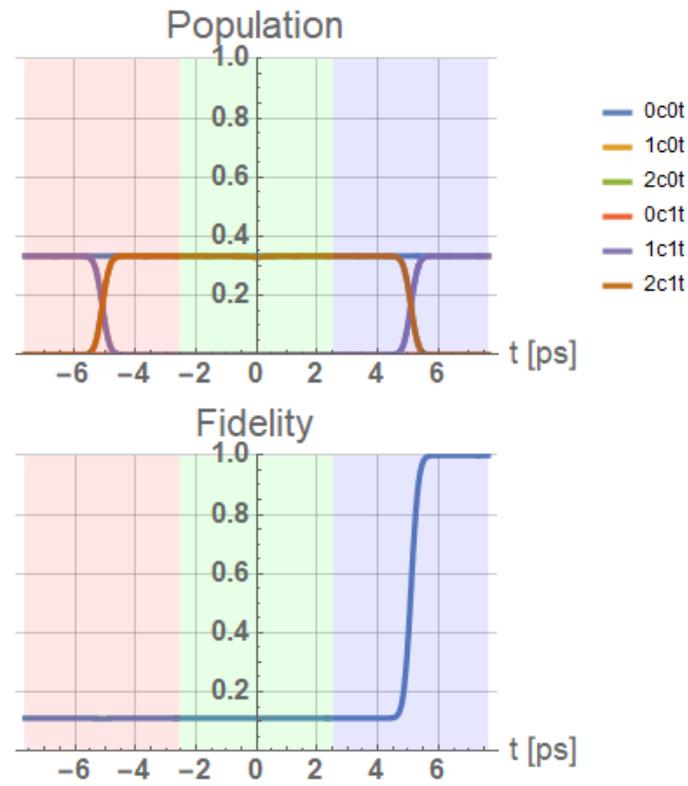

Figure S9 CNOT gate acting on $\frac{1}{\sqrt{3}}(|00\rangle + |10\rangle - |11\rangle)$, the resultant state is $\frac{1}{\sqrt{3}}(|00\rangle - |10\rangle + |11\rangle)$ as shown by the population and fidelity.